\documentclass[aps,pre,nofootinbib,reprint,superscriptaddress]{revtex4-2}

\usepackage[T1]{fontenc}
\usepackage{lmodern}
\usepackage{microtype}
\usepackage[utf8]{inputenc}
\usepackage{graphicx}
\usepackage{subfigure}
\usepackage{xcolor}

\usepackage{placeins}
\usepackage{amssymb}
\usepackage{hyperref}
\hypersetup{
citecolor=red,
colorlinks=true,
filecolor=red,
linkcolor=olive,
linktocpage=true,
urlcolor=cyan
}
\usepackage{enumitem} 

\usepackage{bbm}
\usepackage{amsmath,leftidx}
\usepackage{graphicx}
\usepackage{times}
\usepackage{CJK}
\usepackage{color,colortbl}
\usepackage[normalem]{ulem}

\newcommand{\be}{\begin{eqnarray}}
\newcommand{\ee}{\end{eqnarray}}

\newcommand{\wbe}{\begin{widetext}}
\newcommand{\wee}{\end{widetext}}

\newcommand{\eq}[1]{(\ref{#1})}

\def \nn {\nonumber}

\begin{document}

\title{Can Newtonian Gravity  Produce Quantum Entanglement?}

\author{Feng-Li Lin}
\email{fengli.lin@gmail.com}
\affiliation{Department of Physics, \\
National Taiwan Normal University, Taipei, 11677, Taiwan}

\author{Sayid Mondal}
\email{sayid.mondal@gmail.com}
\affiliation{Instituto de Ciencias Exactas y Naturales, \\ Universidad Arturo Prat, Playa Brava 3256, 1111346, Iquique, Chile}

%================
\begin{abstract}

We investigate whether Newtonian gravity can generate quantum entanglement between mesoscopic quantum bodies modeled as superposed mass quadrupoles using three complementary approaches: mini-superspace, semiclassical gravity, and stochastic gravity. We systematically analyze gravitationally induced entanglement (GIE) mechanisms and the conditions under which they can arise. Our results support the GIE hypothesis by showing that the mini-superspace framework, which quantizes the parity of the gravitational tidal field, can entangle spatially separate quantum bodies. In contrast, the semiclassical and stochastic gravity models, in which the tidal gravitational field sourced by the quantum bodies remains classical, fail to entangle the final state.
These findings clarify recent claims that classical gravity might induce entanglement, and reveal how perturbative treatments can lead to misleading conclusions.

\end{abstract}

\date{\today}

\maketitle

%\tableofcontents
 
%%%%%%%%%%%%%%%%%%%%%%%%%%%%%%%%%%%%%%%%%%%

%\section{Introduction }

\section{Introduction} The unification of quantum mechanics and general relativity remains one of the most profound challenges in fundamental physics. In the absence of direct empirical evidence for quantum gravity at the Planck scale, recent focus has shifted toward low-energy laboratory tests that can witness the non-classicality of the gravitational field. Building on a {\it Gedanken-experiment} originally proposed by Feynman at the 1957 Chapel Hill conference \cite{DeWittRickles2011}, recent table-top proposals \cite{Marletto:2017kzi, Bose:2017nin}—often referred to as the Bose-Marletto-Vedral (BMV) mechanism—aim to verify the quantum nature of gravity\footnote{For several other experimental proposals to realize the GIE protocol, see \cite{Marshman:2019sne,Bose:2022uxe,BeckeringVinckers:2023kco}}. These proposals utilize the dynamics-independent quantum information scheme, known as the gravitationally induced entanglement (GIE) protocol. Based on the local operations and classical communication (LOCC) protocol, the GIE protocol asserts that if gravity can mediate entanglement between two spatially separated massive quantum bodies, the gravitational field itself must possess non-classical features. Consequently, the prevailing consensus within the quantum information framework is that a local classical mediator cannot generate entanglement.

A recent study \cite{Aziz:2025ypo} posits that classical gravity, when coupled with mesoscopic quantum matter, can indeed generate entanglement, despite the classical nature of the mediating field. This claim stands in stark contrast to the dynamical constraints argued in \cite{Marletto:2025asw, Marletto:2025fpm}, which maintain that the detection of gravity-mediated entanglement constitutes a definitive probe of quantum gravity \footnote{See \cite{Nandi:2024jyf} for a theoretical model to explore the quantum nature of gravity.}. The apparent contradiction between the explicit calculations in \cite{Aziz:2025ypo} and the GIE no-go theorems \cite{Marletto:2017kzi, Bose:2017nin} indicates a critical need to bridge the gap between abstract information-theoretic principles and concrete field-theoretic models. While the GIE argument is robust in its generality, its lack of attribution to specific exemplar gravity-matter models leaves it vulnerable to interpretational loopholes arising from technical subtleties in perturbative calculations.

In this Letter, we address this issue by analyzing entanglement dynamics within a rigorous source-theory framework. Rather than relying on generic arguments, we construct a concrete model of gravity-matter interaction to examine the validity of the GIE protocol against the claims of classical-induced entanglement. We define our quantum bodies as mesoscopic superposed states characterized by the $Z_2$ parity of physical observables, specifically modeling them as mass quadrupoles. These mesoscopic quantum bodies for the GIE protocol play an analogous role to the microscopic spin-$1/2$ particles in the Stern-Gerlach experiment.

To systematically test the capacity of gravity to entangle these sources, we generalize the classical Newtonian source theory into three distinct interaction scenarios: (I) the mini-superspace approach to quantum gravity by just quantizing the parity of the gravitational tidal field sourced by the quantum mass quadrupole; (II) the semiclassical gravity approach by sourcing the Newtonian potential by the expectation value of the mass quadrupole; (III) the stochastic gravity approach by appending the semiclassical gravity with a Langevin-type noise due to fluctuation of the parity of quantum mass quadrupole. 
 
By studying the unitarily evolved final state of the two quantum bodies governed by quadrupole-quadrupole interactions in the three scenarios above, we conclude that the quantum bodies can be entangled in scenario (I) but not in scenarios (II) and (III). Therefore, our result is consistent with the GIE protocol. Moreover, in scenarios (II) and (III), we point out a possible loophole based on the perturbative calculation, which will yield an artifact entanglement production due to the neglect of the subleading terms by truncating the expansion at a particular order of the Newton constant. This could be a cause of the disagreement between \cite{Aziz:2025ypo} and the GIE ones \cite{Marletto:2017kzi, Bose:2017nin}.

\section{Preparation of mesoscopic quantum bodies}

Following Feynman's original proposal \cite{DeWittRickles2011}, a massive object may be prepared in a spatial superposition, which we call a "quantum body". Physically, this mass superposition can be realized as distributed N00N states with spatial separation of two different modes  located respectively at $\vec{x}_L$ and $\vec{x}_R$ \cite{Dowling:2008pbf, Kim:2025hbh}:
\be\label{N00N}
|{\rm N00N} \rangle =\frac{1}{\sqrt{2}} \big( |N; \vec{x}_L \rangle |0;\vec{x}_R \rangle + e^{i N \theta} |0; \vec{x}_L  \rangle |N;\vec{x}_R \rangle \big)\;,
\ee
where $N$ denotes the number of photons or atoms, depending on the experimental setup.
Such states have recently been employed to study their quantum decoherence in the presence of a black hole \cite{Danielson:2021egj, Danielson:2022tdw, Danielson:2022sga, Biggs:2024dgp}.

\begin{figure}
    \begin{subfigure}
    \centering
    \includegraphics[width=.9\linewidth]{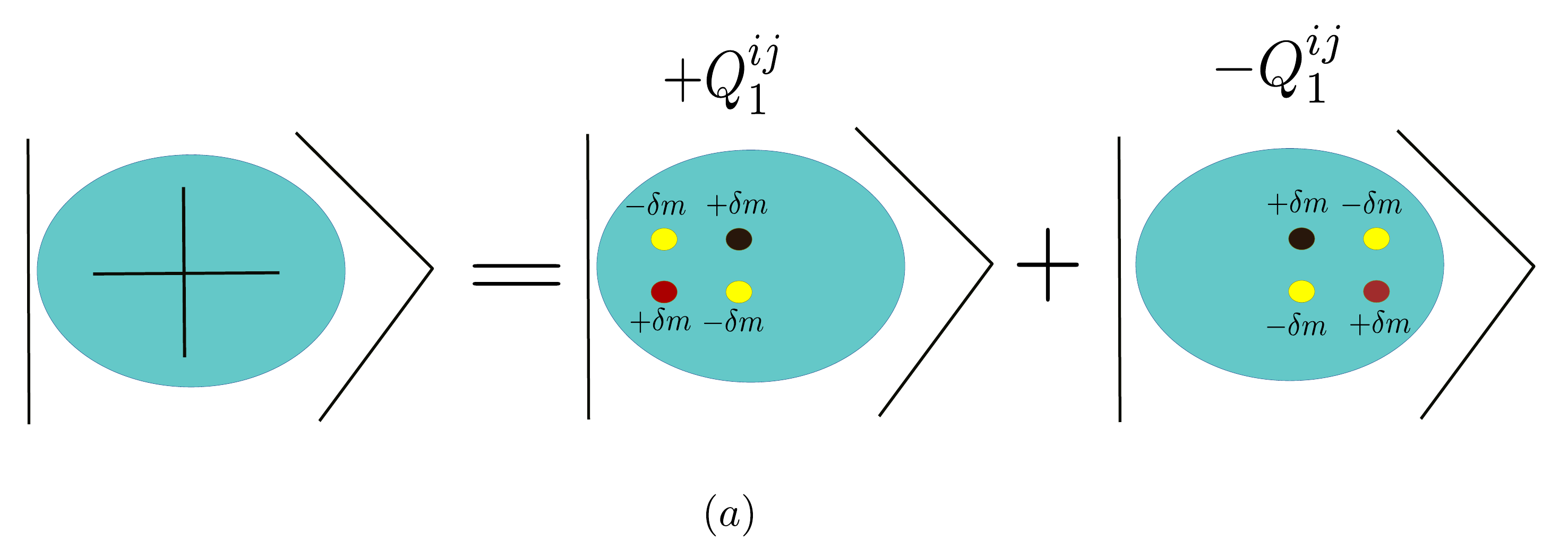}
    %\label{fig1a:main}
    \end{subfigure}
    %\caption{}\label{fig1a}
    \begin{subfigure}
    \centering
	\includegraphics[width=.9\linewidth]{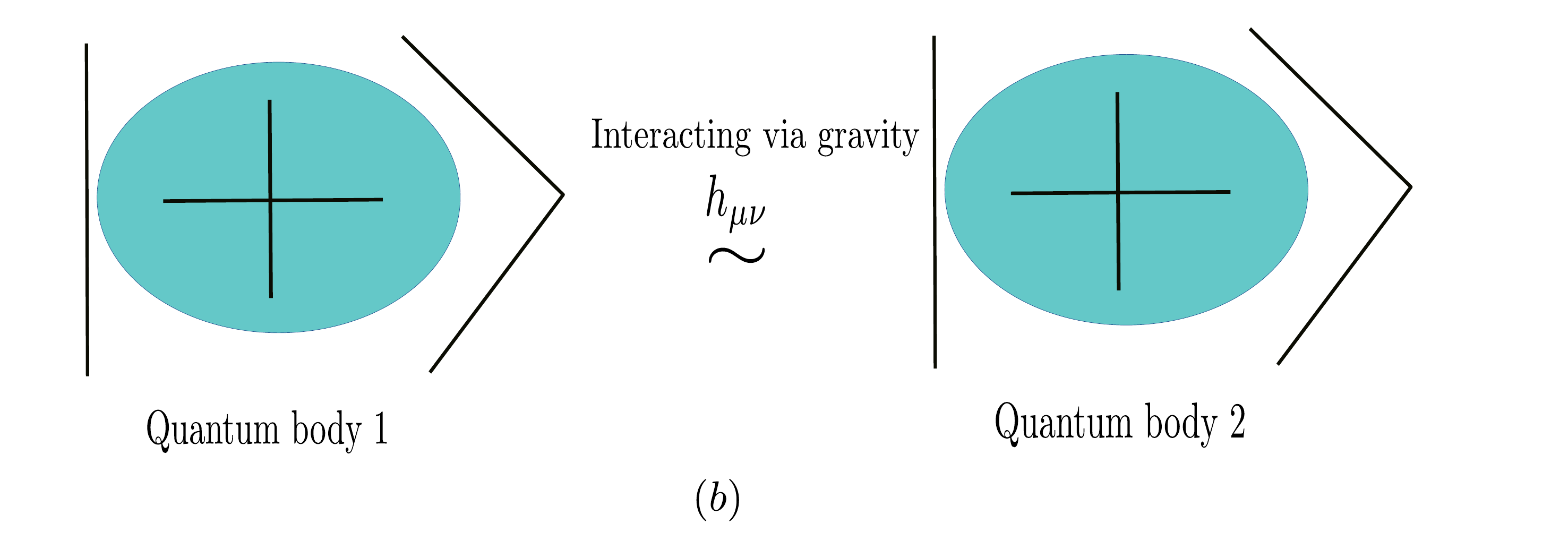}
    %\label{fig1b:main}
    \end{subfigure} 
 \caption{(a) Quantum body of mass quadrupole with its sign in the
equal-weight superposed state. The $\pm \delta m$ are the local mass contrasts of either sign with respect to the underlying uniform mass distribution.  (b) Two such quantum bodies with mass quadrupoles interact via the tidal gravitational fields. The main characteristics of this interaction are dictated by Newtonian gravity; however, its quantum nature will be scenario dependent.} 
\label{fig1:main}
\end{figure}

Moreover, if the N00N state also carries an electric charge, then by placing a charge of opposite sign at $\vec{x}_M =(\vec{x}_L - \vec{x}_R)/2$, we can form a superposed electric dipole. In a gravitational analog—a mass dipole—is forbidden as there is no negative mass. However, it is possible to form the mass quadrupole with its sign in the superposition state as shown in Fig. \ref{fig1:main}(a) by superposing two proper arrangements of the relative positions of two mass dipoles. Consequently, we focus on constructing a quantum mass quadrupole in which the sign of the quadrupole moment is in superposition, and study the interactions of two such quantum bodies as shown in Fig. \ref{fig1:main}(b).

Formally, we can associate with such a quadrupole observable a mini-superspace operator $\hat{Q}_{ij}= Q_{ij} \hat{Z}$, where $Q_{ij}$ is a specific classical mass quadrupole,  and $\hat{Z}$ is a $\mathbb{Z}_2$ operator for measuring the sign of the quadrupole, satisfying $\hat{Z}|\pm; Q_{ij}\rangle = \pm |\pm; Q_{ij}\rangle $
with $\langle \pm; Q_{ij}|\pm ; Q_{ij}\rangle =1$ and $\langle \mp; Q_{ij}|\pm; Q_{ij}\rangle =0$.
The $\hat{Q}_{ij}$'s eigenstates  $|\pm; Q_{ij}\rangle$'s correspond to the classical quadrupole moments with definite signs. We can then define a general superposed state of quadrupole signs as follows,
\be\label{theta_phi_s}
|\theta,\phi; Q_{ij}\rangle := \cos\theta |+; Q_{ij}\rangle + \sin\theta e^{i\phi} |-;Q_{ij}\rangle\;.
\ee
The N00N state \eq{N00N} corresponds to the specific case where $\theta=\pm {\pi \over 4}$, $\phi=0$. The expectation value is 
\be
\langle \theta, \phi; Q_{ij}|\hat{Q}_{ij}| \theta,\phi; Q_{ij}\rangle &=& Q_{ij} \langle \theta, \phi; Q_{ij}|\hat{Z}| \theta,\phi; Q_{ij}\rangle \;,\nn\\
&=& Q_{ij} \cos 2\theta\;.
\ee
 
This mini-superspace approach to quantum bodies can be thought of as the mesoscopic version of microscopic quantum spin-${1\over 2}$. Its simplicity helps elucidate the issue discussed in this Letter, much as the role of the quantum spin-${1 \over 2}$ in Stern-Gerlach experiments demonstrates quantum superposition.

\section{Newtonian Gravity of Quantum Bodies}

 In Newtonian gravity, the Newton potential $\Phi$ in the multipole expansion takes the form 
\be\label{Phi}
\Phi(\vec{x}) = G_N\left({ m\over r} + {p_i x_i \over r^3} + { Q_{ij} x_i x_j \over r^5} + \cdots\right)\;,
\ee
where $G_N$ is the Newton constant, $m$, $p_i$, and $Q_{ij}$ are respectively the mass moments of monopole, dipole, and quadrupole of a compact body.  This expansion is valid whenever the field distance $r$ is far larger than the size of the mass source, such as quantum bodies, which is the situation assumed in this Letter. Since there is no negative mass, the mass dipole can be made to vanish by placing the origin at the center of mass. Moreover, the quadrupole moments are symmetric and traceless, i.e., $Q_{ij}=Q_{ji}$ and $\sum_{i=1}^3 Q_{ii}=0$.

The interaction Hamiltonian of a compact body with the Newtonian gravity in the multipole expansion takes the form
\be\label{H_N}
H_{\rm N} = - m \Phi\vert_{\vec{x}=\vec{0}} - p_i \partial_i \Phi\vert_{\vec{x}=\vec{0}} - Q_{ij} \partial_i \partial_j \Phi\vert_{\vec{x}=\vec{0}} + \cdots \;.
\ee
Combining \eq{Phi} and \eq{H_N}, we can obtain the interaction between two static quadrupole moments $Q_{ij}^{(1,2)}$ depicted in Fig.\ref{fig1:main}(b) with the superscript labeling the quantum bodies,
\be\label{HQQc}
H_{QQ}^c &=& -{1\over 2} \Big[ Q^{(1)}_{ij} \partial_i \partial_j \Phi^{(2)}\big\vert_{\vec{x}=\vec{x}_2}  + Q^{(2)}_{ij} \partial_i \partial_j \Phi^{(1)}\big\vert_{\vec{x}=\vec{x}_1} \Big]\;,\nn\\ &=&  -{G_N \over r^5_{12}} Q_{ij}^{(1)} T_{ijkl} Q^{(2)}_{kl} \;,
\ee
where
\begin{widetext}
\be
T_{ijkl} := r^5 \partial_i \partial_j \big[{x_k x_l \over r^5} \big]\Big\vert_{\vec{x}=r_{12}\hat{r}} = \big[\delta_{ik}\delta_{jl} + (i\rightarrow j) \big] - 5\big[ \delta_{jk} \hat{r}_i \hat{r}_l +  \delta_{il} \hat{r}_j \hat{r}_k + (i\rightarrow j) \big] - 5 \delta_{ij} x_k x_l + 35 x_i x_j x_k x_l\;,
\ee
\end{widetext}
with $\hat{r} := { \vec{x}_1 - \vec{x}_2 \over |\vec{x}_1 - \vec{x}_2|}$ and $r_{12}:=|\vec{x}_1 - \vec{x}_2|$. This is the tidal interaction between two static and well-separated quadrupole moments, and can also be reproduced from the leading-order post-Newtonian approximation of Einstein gravity \cite{Binnington:2009bb, Kol:2011vg, Biggs:2024dgp, Lin:2024bmg}, with a possibly different $T_{ijkl}$ due to the choice of gauge fixing.  Moreover, the subleading PN terms will involve the motion of the compact bodies \cite{Goldberger:2007hy, Porto:2016pyg}, which we will not consider in this note.

We now aim to generalize the above source theory of Newtonian gravity for classical bodies to quantum bodies in the superposed states of the quadrupole sign. Due to the lack of a comprehensive, consistent approach to quantum gravity, there are three possible scenarios depending on how to treat the gravity field sourced by quantum bodies. 

\subsection{Mini-superspace approach}

We first adopt a mini-superspace approximation to quantum gravity, wherein the induced Newtonian potential acquires a quantum character via the quadrupole sign operator:
\be
\hat{\Phi}^{{\rm mini},(a)}_{\rm quad} = {G_N Q^{(a)}_{ij} x_i x_j \over r^5} \hat{Z}^{(a)}, \qquad a=1,2\;,
\ee
where $\hat{Z}^{(a)}$ represents the $\mathbb{Z}_2$ parity operator for the respective quantum bodies.
While this may not be a consistent treatment of quantum gravity, it is conceptually analogous 
 to the mini-superspace approach in quantum cosmology \cite{DeWitt:1967yk, Hartle:1983ai, Vilenkin:1984wp}, which involves quantizing only the scale factor, not the full metric.

Generalizing \eq{HQQc} to this framework, the Hamiltonian for the interaction between two such quantum bodies reads
\be\label{Hq_1}
\hat{H}_{QQ}^{\rm mini} = -\hat{Q}^{(1)}_{ij} \partial_i\partial_j \hat{\Phi}^{{\rm mini},(2)}_{\rm quad} \big\vert_{\vec{x}=\vec{x}_2} = H_{QQ}^c \hat{Z}^{(1)} \otimes \hat{Z}^{(2)}
\ee
with  $H_{QQ}^c$ is defined in \eq{HQQc}. 

\subsection{Semi-classical gravity approach}

In the semi-classical gravity framework~\cite{dewitt1975quantum,parker1968particle,Hawking:1975vcx}, matter is quantized while the gravitational field remains classical. Consequently, the Newtonian potential is sourced by the expectation value of the quadrupole operator with respect to the N00N-like state, rather than the operator itself:
\begin{eqnarray}\label{Phi_semi}
\Phi^{{\rm semi},(a)}_{\rm quad} &=& \frac{G_N x_i x_j}{r^5} \langle \theta^{(a)}, \phi^{(a)}|\hat{Q}^{(a)}_{ij}| \theta^{(a)}, \phi^{(a)} \rangle \;,\nonumber\\
&=& \frac{G_N Q^{(a)}_{ij} x_i x_j}{r^5} \cos2\theta^{(a)}\;, \qquad a=1,2\;.  
\end{eqnarray}

Even though the Newton potential sourced by a quantum body is a classical quantity, this hybrid system exhibits bizarre behavior when a measurement collapses the quantum body, e.g., from $\theta={\pi \over 4}$ to $\theta=0$. This will then result in a sudden jump of the Newton potential due to the wavefunction collapse \cite{Penrose:1996cv}. In general relativity, this implies that the spacetime metric can be singular.

Given the classical nature of the source in (\ref{Phi_semi}), the interaction Hamiltonian takes the form:
\begin{widetext}
\begin{equation}\label{HQQc_semi}
\hat{H}_{QQ}^{\rm semi}= \frac{1}{2} H_{QQ}^c \left[ (I^{(1)} \otimes \hat{Z}^{(2)}) \cos 2\theta^{(1)} + (\hat{Z}^{(1)} \otimes I^{(2)}) \cos 2\theta^{(2)}\right]\;,
\end{equation}
\end{widetext}
where $I^{(a)}$ denotes the identity operator for the $a$-th body. Crucially, the quantum nature of $\hat{H}_{QQ}^{\rm semi}$ arises solely from the probe body, not the source field. For the specific N00N state in (\ref{N00N}), $\Phi^{{\rm semi},(a)}_{\rm quad}$ vanishes as $\theta^{(a)}=\pi/4$.

\subsection{Stochastic gravity approach}
Stochastic gravity extends the semi-classical approximation by incorporating fluctuations in the matter stress-energy tensor via the Einstein-Langevin equation~\cite{Hu:1994dka, Hu:2008rga, Martin:1998nc}:
\begin{equation}
G_{\mu\nu} = 8 \pi G_N \big(\langle T_{\mu\nu} \rangle + \xi_{\mu\nu} \big)\;.
\end{equation}
Here, the gravitational field is driven by a Gaussian stochastic source $\xi_{\mu\nu}$, arising from the noise kernel (fluctuations of the quantum matter)~\cite{Hu:1994dka}. This is the generalization of Langevin dynamics for Brownian motion, with the gravity field as the Brownian particle and the quantum matter as the thermally-agitated water molecules.   At the Newtonian order, the potential in (\ref{Phi_semi}) acquires a random component corresponding to fluctuations in the quadrupole orientation. This is modeled by modifying the orientation parameter $\theta^{(a)}$:
\begin{equation}
\theta^{(a)} \rightarrow \tilde{\theta}^{(a)} = \theta^{(a)} + \delta\theta\;,
\end{equation}
where $\delta\theta$ represents Gaussian white noise $\mathcal{N}(0,1)$. Consequently, the stochastic interaction Hamiltonian $\hat{H}_{QQ}^{\rm stoc}$ retains the form of Eq.~(\ref{HQQc_semi}), with $\theta^{(a)}$ replaced by the stochastic variable $\tilde{\theta}^{(a)}$.

\section{Does Newtonian gravity entangle quantum bodies?}
With the descriptions of three different source theories for quantum bodies and the corresponding Newtonian quadrupole-quadrupole interaction Hamiltonians, we are now ready to examine whether these Newtonian interactions can entangle two spatially separated quantum bodies.  This is basically to examine the evolved state
\be
|{\rm final} \rangle = U_T  |{\rm init} \rangle 
\ee
with the unitary evolution operator 
\be
U_T := {\cal T} \exp\left({-i \int_{-\infty}^{\infty} w_T(t) \hat{H}_{QQ} dt}\right) \;,
\ee
where $\hat{H}_{QQ}$ corresponds to either the mini-superspace ($H_{QQ}^{\rm mini}$), semiclassical ($H_{QQ}^{\rm semi}$), or stochastic ($H_{QQ}^{\rm stoc}$) Hamiltonian, and ${\cal T}$ is the time-odering operator. The window function is chosen as $w_T(t)= e^{-(t/T)^2}(\pi/2)^{-1/4}$, satisfying the normalization $\int |w_T|^2 dt =T$. We choose the initial state of the two quantum bodies to be a product state
\be
|{\rm init} \rangle = |\theta^{(1)},\phi^{(1)}; Q^{(1)}_{ij} \rangle \otimes |\theta^{(2)},\phi^{(2)}; Q^{(2)}_{ij} \rangle
\ee
requiring $\theta^{(a)}\ne n \pi/2$ ($n \in \mathbb{Z}$) to ensure the bodies are in macroscopic superpositions rather than eigenstates of the interaction basis, otherwise it would mimic classical behavior.

\subsection{Entanglement production in mini-superspace approach}
For the mini-superspace approach with $\hat{H}_{QQ}^{\rm mini}$ is defined in \eq{HQQc_semi}, then the time evolution operators simplifies to
\be
U_T = e^{-i \lambda \hat{Z}^{(1)} \otimes \hat{Z}^{(2)}} 
\ee
with

\be
\lambda &:=& - H_{QQ}^c \int_{-\infty}^{\infty} dt \; w_T(t)\;= - (2\pi)^{1/4} H_{QQ}^c T \;,\\
&\sim &{\cal O}(1) {G \delta M_1 \delta M_2 \over r_{12}} \Big({b\over r_{12}}\Big)^4 T\;.
\ee

For simplicity, we consider a specific initial state with $\theta^{(1)}=\theta^{(2)}={\pi\over 4}$ and $\phi^{(1)}=\phi^{(2)}=0$, i.e.,
\be\label{init_0}
|{\rm init} \rangle = |++\rangle := |+; 1\rangle \otimes |+; 2\rangle\;, 
\ee
where from \eq{theta_phi_s} we have defined 
\be
|+;a \rangle := |\theta^{(a)}=\pi/4, \phi^{(a)}=0; Q^{(a)}_{ij}\rangle 
\ee
with $a=1,2$ labeling the quantum bodies. To avoid notational clumsiness, here and below we use $|\pm; a\rangle$ to denote the eigenstates of Pauli $\hat{X}^{(a)}$ operators, and one should not confuse them with the $|\pm; Q^{(a)}_{ij}\rangle$ used in \eq{theta_phi_s} for the eigenstates of $\hat{Z}^{(a)}$.  

Furthermore, we define
\be
|--\rangle := |-; 1\rangle \otimes |-; 2\rangle \;, 
\ee
where
\be
|-; a\rangle := \hat{Z}^{(a)}|+; a\rangle \quad {\rm and} \quad  \langle \pm; a|\mp;a\rangle = 0\;.
\ee
Similarly, one can define $| +- \rangle$ and $|-+ \rangle$. 
Then, the time-evolved final states become
\be
|{\rm final}\rangle= U_T |++\rangle =\cos\lambda \; |++\rangle - i \sin\lambda \; |-- \rangle\;.
\ee
For $\lambda \neq n\pi/2$ where $n\in \mathbb{Z}$, the Schmidt rank increases to two, confirming that the mini-superspace formulation—representing quantized gravity—entangles the bodies. This aligns with the GIE protocol, establishing entanglement generation as a probe for quantum gravity.

\bigskip

\subsection{No entanglement production in semiclassical and stochastic gravity approaches}
Now, for the semiclassical gravity and stochastic gravity, the $\hat{H}_{QQ}$ takes the form of \eq{HQQc_semi}. In such cases, $U_T$ can be factorized into the product of two local unitaries, i.e.,
\be
U_T = e^{-\lambda_2 \hat{Z}^{(1)}} \otimes e^{-\lambda_1 \hat{Z}^{(2)}}\;,
\ee
where
\be\label{lambda_a}
\lambda_a = -{1\over 2} (2\pi)^{1/4} H_{QQ}^c T \cos 2 \Theta^{(a)}\;, \qquad a=1,2
\ee
with $\Theta^{(a)}=\theta^{(a)}$ for semiclassical gravity, and $\tilde{\theta}^{(a)}$ for stochastic gravity. The product of local unitaries cannot produce entanglement, as can be seen as follows,
\be\label{no_ent}
|{\rm final} \rangle = && e^{-\lambda_2 \hat{Z}^{(1)}} |\theta^{(1)},\phi^{(1)}; Q^{(1)}_{ij} \rangle
 \nn \\
 && \otimes \; e^{-\lambda_1 \hat{Z}^{(2)}} |\theta^{(2)},\phi^{(2)}; Q^{(2)}_{ij} \rangle\;,   
\ee
which remains as a product state. This is consistent with the GIE proposal again, i.e., the classical gravity cannot produce quantum entanglement.

\subsection{Confusing GIE-violating result due to perturbative calculations}
The recent paper \cite{Aziz:2025ypo} claims that classical gravity can produce quantum entanglement, which is based on perturbative calculations in the context of the stochastic gravity approach. We now demonstrate a subtlety for reaching this GIE-violating result. For simplicity, we consider the initial state of \eq{init_0} so that $\lambda_a$'s of \eq{lambda_a} vanish as $\cos 2\theta^{(a)}\big\vert_{\theta^{(a)}=\pi/4}=0$ for semiclassical gravity approach, but not for stochastic gravity approach due to nontrivial $\delta \theta$. So, there is no evolution for the semiclassical gravity in this case. On the other hand, for the stochastic gravity, we have
\be\label{no_ent_1}
|{\rm final} \rangle = e^{-\lambda_2 \hat{Z}^{(1)}} |+;1\rangle \otimes e^{-\lambda_1 \hat{Z}^{(2)}} |+;2 \rangle\;,
\ee
which is unentangled. 

Now, we can explain the possible origin of claiming the GIE-violating result as follows. In the perturbative approach as performed in \cite{Aziz:2025ypo}, the resultant final state is obtained up to some order of $G_N$. This corresponds to expanding the right side of \eq{no_ent_1} in order of $\lambda_a$ as  $\lambda_a \sim {\cal O}(G_N)$. Suppose we are interested in the final state up to ${\cal O}(G_N)$, then we have
\begin{widetext}
\be
|{\rm final}\rangle & =& \big(1-i\lambda_2 \hat{Z}^{(1)} + \cdots \big) |+;1\rangle \otimes \big(1-i\lambda_2 \hat{Z}^{(2)} + \cdots \big) |+;2\rangle\;,\nn\\
& \simeq &|++\rangle - i\lambda_1 |+-\rangle -i \lambda_2 |-+\rangle + {\cal O}(G_N^2)\;.  
\ee
\end{widetext}
We see that the above final state up to ${\cal O}(G_N)$ is now entangled due to the neglect of the $\lambda_1 \lambda_2 \sim {\cal O}(G_N^2)$ term. The entanglement entropy of the final state is
\be
S\sim \lambda_1^2 \lambda_2^2 \sim {\cal O}(G_N^4)\;.
\ee
This is tiny but still non-zero even after averaging over $\delta\theta$ as $\overline{\lambda_a^2}\sim \overline{\delta \theta \delta \theta} \sim \sigma_{\delta \theta}^2$ with overline denoting the average over Gaussian noise and $\sigma_{\delta \theta}$ its variance. However, the entanglement induced in this way is just an artifact due to the truncation of higher-order terms in the perturbative calculation. 

\section{Discussion and conclusion}
 In this article, we systematically investigated the capacity of Newtonian gravity to induce quantum entanglement between mesoscopic quantum bodies, specifically modeled as mass quadrupoles. Our theoretical framework analyzed three distinct interaction scenarios to elucidate the mechanisms of entanglement generation under gravitational influence: (1)  mini-superspace approach, (2) semiclassical gravity, and (3) stochastic gravity.

Our results provide robust support for the gravity-induced entanglement (GIE) hypothesis, reinforcing the postulate that classical gravity cannot mediate quantum entanglement. We demonstrated that the mini-superspace approach successfully facilitates entanglement production. This is achieved by quantizing the parity of gravitational tidal fields sourced by the mass quadrupoles. Conversely, both semiclassical and stochastic gravity models failed to generate entanglement, yielding only final product states. These findings confirm the necessity of a quantum gravitational framework for the effective production of entanglement, aligning with the core implications of the GIE protocol.

A critical aspect of our work involves addressing recent assertions that classical gravity might induce entanglement under specific conditions. We clarified that such claims likely arise from perturbative calculations that neglect higher-order cross-terms. By isolating these artifacts, we demonstrated how truncations in expansions can lead to misleading interpretations, thereby resolving potential confusion regarding the quantum nature of gravitational interactions.

Our results establish a foundation for further exploration into the interplay between gravity and quantum systems. As experimental techniques continue to advance, developing concrete models grounded in rigorous theoretical frameworks will be essential to probing the intricate link between gravity and quantum mechanics, ultimately deepening our understanding of fundamental physics.

\vspace{2 cm}

\section{Acknowledgements}

 SM is grateful to Ignacio Araya and Giorgos Anastasiou for their hospitality at Universidad Andrés Bello and Universidad Adolfo Ibañez, respectively. The work of FLL is supported by Taiwan's National Science and Technology Council (NSTC) through Grant No.~112-2112-M-003-006-MY3. The work of SM is supported by the ANID FONDECYT Postdoctorado grant number 3240055.

%\bibliographystyle{apsrev4-1}
%\bibliography{refs}

\end{document}